\documentclass{webofc}
\usepackage[varg]{txfonts}   
\usepackage{amsmath,amsfonts,verbatim,graphicx,float,epsfig,rotating,appendix}
\begin{document}
\title{Cosmological Constant Effects on the Properties of Mass Twin Compact Stars}
%
%

\author{\firstname{Noshad} \lastname{Khosravi Largani}\inst{1}\fnsep\thanks{\email{egradaat@gmail.com}} \and
        \firstname{David Edwin} \lastname{\'Alvarez-Castillo}\inst{2}\fnsep\thanks{\email{alvarez@theor.jinr.ru}}
}

\institute{Department of Physics, Alzahra University, Tehran, 1993893973, Iran 
\and
           Bogoliubov  Laboratory of Theoretical Physics, JINR Dubna, 141980 Dubna, Russia
          }

\abstract{%
  
We present a systematic investigation of the cosmological constant effects in compact stars interiors in the framework of Einstein's gravity.
Consideration of a cosmological constant $\Lambda$ in compact stars is motivated by the mechanism of acceleration of the observable universe,
where $\Lambda$ is usually related to the dark energy. In particular, we consider compact star mass twins, hybrid neutron stars that 
populate both the second and third branch of the mass-radius diagram. For those models, 
the need of consideration of excluded volume effects in the equation of state, resulting from the finite size volume of nucleons, leads to a stiffening
of matter causing compact stars to acquire higher mass and radius values. We demonstrate that certain values of the cosmological constant can also modify 
the compact star properties but in an opposite way. In addition, we find that the inclusion of $\Lambda$ can have a similar effect to the existence of pasta phases at the hadron-quark interface.
}
\maketitle
\section{Introduction}

The physics of neutron stars is currently a very active topic of investigation, covering many aspects of the contemporary physics. These include, but not only, astrophysical processes like energetic emissions,
stellar evolution, studying the dense matter properties in compact star interiors, and element nucleosynthesis. The recent detection of gravitational radiation from the event GW170817~\cite{TheLIGOScientific:2017qsa} helped to clarify many of the questions related to the aforementioned phenomena, like corroborating that the only possibility of creation of certain heavy elements is through the fusion of neutron stars and its association with kilonovae, as well as determination of masses of the binary system. On the other hand, Cosmological studies are dedicated to the understanding of the dynamics of universe as a whole as well as considering the role of the fundamental interactions within it. In this respect, $\Lambda$ is one of the necessary ingredients for the description of the observable universe, providing a mechanism for its accelerated expansion, somehow related to the so called dark energy. 

In this work, we take neutron stars as a probe of the possible effects of the cosmological constant as a free parameter, and allow it to vary inside the star, where matter effects might change its strength, in contrast to the conditions existing  in vacuum. We consider two up-to-date realistic equations of state  (EoS) and look at the resulting macroscopic changes in the stars, namely mass and radius. Of these two chosen EoS, the second features two mass twin stars located in the second and third branch in the mass-radius diagram and brings the possibility of probing a critical point in the QCD phase diagram~\cite{Blaschke:2013ana}. The approach in this paper is indeed motivated by the fact that whenever one faces a new phenomenon that cannot be explained with the already existing theories, two evident possibilities exist: a) assuming that there is a completely new theory which has not been developed yet, or b) trying to modify the current valid paradigm in order to broaden the domain of exploration. Our $\Lambda$ parameter in this study might account for vacuum energy effects~\cite{Csaki:2018fls} as well as to correspond to some alternative gravity theories parameter under certain limits, thus providing a useful benchmark.

\section{The role of the cosmological constant in today's cosmological picture}

Although it seems that Einstein's General Relativity is working perfectly well in our solar system, some problems including a reason for accelerated expansion of the universe and the lack of precise data beyond this scale has made some physicists to think out the box and to consider alternative theories of gravity such as Scalar-Tensor theories, $f(R)$ models, to name a few. 
Moreover, many approaches have started to modify many of the different aspects of the cosmological standard model, the so called $\Lambda$CDM. Nevertheless, the recent GW170817 event detection of simultaneous electromagnetic and gravitational signals was able to rule out many of these possibilities~\cite{Boran:2017rdn}.

\section{Compacts stars under modified gravity approaches}

The standard way of modelling neutron stars starts by considering a static, spherical and not rotating stellar object under a given theory of gravity which will require interior and external solutions matching properly at the boundaries. In the case of general relativity, this treatment leads to the Tolman-Volkov-Oppenheimer equations (TOV)~\cite{Tolman:1939jz,Oppenheimer:1939ne} (here written in the natural units $c=G=1$):
\begin{equation}
 \frac{dm}{dr}= 4\pi r^2\epsilon
 \label{TOVmass}
\end{equation}
\begin{equation}
 \frac{dP}{dr}=-\frac{(\epsilon+ {P})(m+4\pi r^3 {P})}{r(r-2m)}
  \label{TOVpressure}
\end{equation}
where integration of the first equations will determine the total mass $M$ of the star, where as the second one will allow for determination of the pressure profile as a function of radius as well as the total stellar radius $R$ when $P(r=R)=0$. This equations are to be solved by choosing a central density $\epsilon_c$ for a given star and increasing it to obtain another more massive up to the so called maximum mass $M_{Max}$, the last configuration stable against radial oscillations produced when gravity takes over.
 Most of alternative theories of gravity attempts of description of compact stars end up with modifications to these equations. Those modifications will modify the mass-radius diagram by increasing or decreasing either values, like changing the maximum mass or minimal radius.  For the most general free variation of parameters that can be included the TOV equations see~\cite{Velten:2016bdk}, where a classification of the corresponding modifications for most popular alternative gravity theories can be found. 

\section{The compact star equation of state}

In order to test the effect of the cosmological constant parameter we fix the equation of state. We utilise two realistic EoS that fulfil modern constraints from terrestrial laboratory experiments as well as astrophysical observations. In one hand, we select the DD2 relativistic density functional model EoS~\cite{Typel:2009sy} which is used to describe a classical neutron star whose interior contains only protons and neutrons and leptons like electrons and possibly muons as well. Its great quality relies on the fact that it has been calibrated to nuclear physics measurements of properties like parameter values at saturation as well as those of the symmetry energy.

In addition, we consider in our study a multi-polytrope (MP) EoS whose parameters were fit to realistic EoS in the lower density regions and features a strong first order phase transition at a high density value which can represent the transition from hadron to quark matter inside compact stars. This EoS model has been introduced in~\cite{Alvarez-Castillo:2017qki} where the authors discuss the \textit{high mass twins} phenomenon (HMTs) where stars with very similar masses are located in different branches of the mass-radius diagram therefore having considerable
different radii ($\approx1-2$km difference).  This case is of two-fold relevance since it serves to solve several microscopic issues like the reconfinement, masquerades, and the hyperon puzzle (see~\cite{Blaschke:2015uva} for an extended discussion) as well as providing mass and radius predictions potentially observable by nowadays facilities like the NICER X-ray detector~\cite{Arzoumanian:2009qn} and gravitational wave observatories that can estimate tidal deformabilities that strongly related to those stellar properties~\cite{Hinderer:2009ca, Paschalidis:2017qmb,Alvarez-Castillo:2018pve}.  

\section{Results}
\begin{figure*}[!bhtp]
\begin{center}$
\begin{array}{cc}
\includegraphics[width=0.60\textwidth]{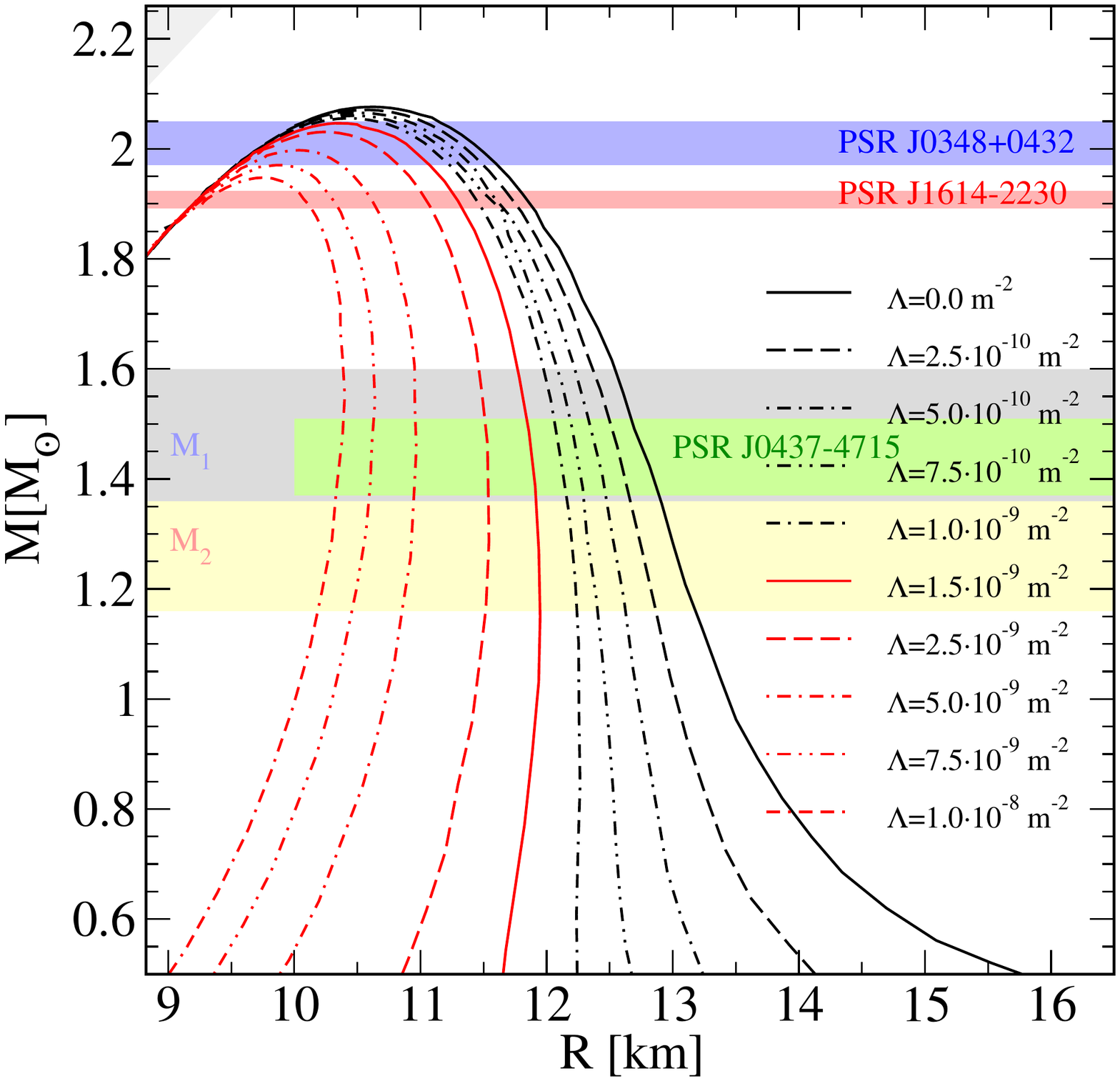} &\hspace{-2cm} \includegraphics[width=0.60\textwidth]{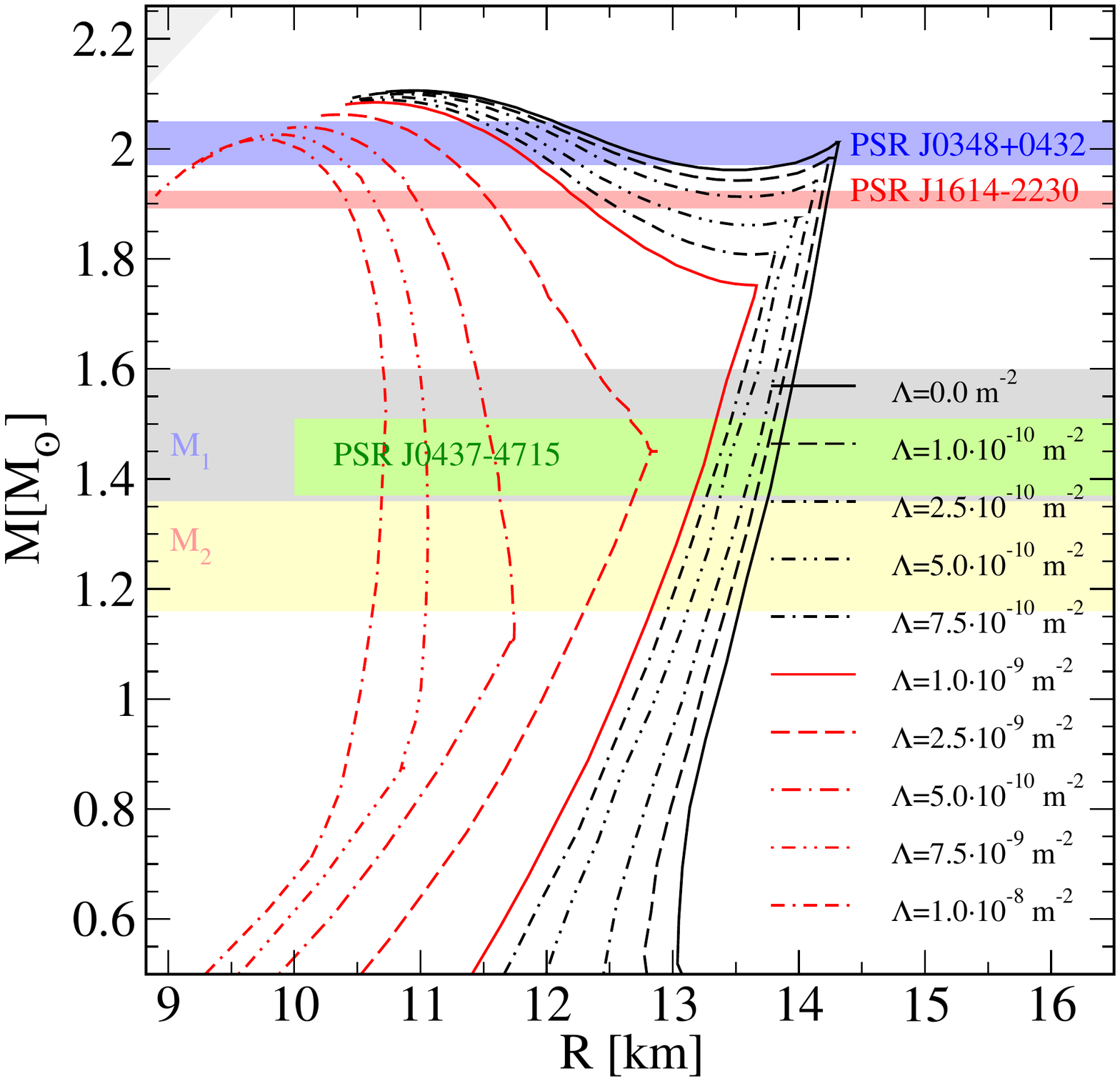}
\end{array}$
\end{center}
\caption{\label{MvsR}
Mass-Radius for compact stars under the influence of a cosmological constant $\Lambda$. The shaded colourful areas correspond to measurements of neutron stars, see~\cite{Alvarez-Castillo:2018pve} for details.  Left column: DD2 EoS. Cosmological parameter values can shrink mass and radius values, eventually changing the topology from a standard hadronic EoS (black lines) into a quark EoS (red lines) in the framework of general relativity with zero cosmological constant.  
Right column: High mass twins (MP-HMTs) EoS. The strength of the $\Lambda$ parameter can wipe out the mass twins (black lines) leaving a quark-like EoS (red lines) with a characteristic kink due to the strong first order phase transition in 
the HMTs EoS. This phenomenon is similar to an internal softening of the EoS when a mixed phase featuring geometrical structures appear~\cite{Alvarez-Castillo:2017xvu,Ayriyan:2017nby}, however no kink is present in that case.} 
\end{figure*}

In this section we present results of our investigation on the effects of the cosmological constant inside compact stars. We start by considering the same general relativity TOV of equations with the inclusion of a $\Lambda$ parameter~\cite{2014AN....335..593Z,Bordbar:2015wva}, where the mass equation~(\ref{TOVmass}) remains the same but the one for the pressure~(\ref{TOVpressure}) is modified as:
\begin{equation}
 \frac{dP}{dr}=-\frac{(\epsilon+ P)(m+4\pi r^3 P+\Lambda r^{3}/3)}{r(r-2m+\Lambda r^{3}/3)}.
\end{equation}
It is then clear that the term $\Lambda r^{3}/3$ is the cosmological contribution to two of the factors in the TOV equations. Therefore, all the modifications to the compact star properties effectively modify the internal pressure profiles of stars. These new equations are solved under the same initial and boundary conditions as in the zero cosmological constant case. The resulting mass-radius diagrams for both equations of state considered here are presented in figure~\ref{MvsR}, where the left panel corresponds to the pure hadronic DD2 EoS and the right panel to the MP-HMTs EoS. 

Within our figures, we also present shaded regions that correspond to mass estimates for several objects. Such values are often derived from measurements of other quantities therefore in order to extract the corresponding mass values, general relativity is applied to the modelling of the involved physical process. Examples include binary systems dynamics, gravitational redshifts, gravitational wave emission and tidal deformabilities, among others. With that being said, it is necessary to include the modifications to gravity into the mass estimation from the measured quantities. Nevertheless, we include the standard mass measurements in our mass-radius plots as a reference. 
We summarise our findings in the following section.
\section{Outlook and conclusions}

In this paper we have investigated changes that can be possibly occur in the compact star morphology by considering the variation of a parameter that we associate with the cosmological constant $\Lambda$.  The main motivation is to explore possible effects of $\Lambda$ with matter inside neutron stars, where the quantum vacuum might behave differently than in the cosmic voids. In the other hand, alternative theories of gravity often modify the compact star structure as well. For instance, the inflaton model presented in~\cite{Fiziev:2016vpz} features a scalar field that modifies the internal stellar pressure. Other examples include chameleons fields~\cite{Chagoya:2018lmv} and $f(R)$ theories~\cite{Folomeev:2018ioy}, as well as more involved approaches like Palatini gravity~\cite{Wojnar:2017tmy}.
In our endeavour for exploring related effects we have chosen two state-of-the-art equations of state and solved the general relativistic equations with cosmological constant as presented in~\cite{Bordbar:2015wva,Hendi:2015vta}. We have found
that the typical values for $\Lambda$ to have an observable effect in compact stars are of the order of $\Lambda=10^{-8} - 10^{-10} $ m$^{-2}$. Our first observation is that the  $\Lambda$ values we obtain differ in up to five orders of magnitud from those of~\cite{Bordbar:2015wva}, apparently due to the stiffness of our EoS  sets in contrast to the soft EoS presented there. The stiffness in our models is indeed an important requirement to support massive compact stars, like the most massive measured one of 2M$_{\odot}$~\cite{Antoniadis:2013pzd}.
Moreover, we have found that the strength of $\Lambda$ can change the topology of mass-radius curve, see figure~\ref{MvsR}. The general tendency is that higher values of cosmological constant tend to lower the maximum mass and shrink the radius. This effect is quite the opposite to the rise of pressure from the microscopic description of the equation of state when for instance an excluded volume of the nucleons produced by the Pauli blocking of their internal quarks is considered~\cite{Benic:2014jia,Alvarez-Castillo:2016oln}.  As $\Lambda$ is increased pure hadronic stars change their appearance from what is the standard hadronic mass-radius curve to the typical one of a quark star, an object composed of a plasma of quarks bound only by gravity.  Furthermore, this effect is also present in the HMTs stars, with the $\Lambda$ highest values causing the instabilities in the mass-radius diagram to disappear thus frustrating the existence of mass twins. A similar mass-radius behaviour has been observed when the EoS is modified by considering pasta phases at the hadron-quark phase transition~\cite{Alvarez-Castillo:2017xvu,Ayriyan:2017nby}. However, for the cosmological parameter case, there remains a sharp kink in the mass-radius relation that evidences the nature of its EoS and the strong first order phase transition in it.

\section{Acknowledgments}

The authors would like to express their gratitude to the organisers of the AYSS 2018 conference for their hospitality and support. D.E.A-C. acknowledges support from the Ter-Antonyan-Smorodinsky program for collaboration between Armenian Institutions and JINR.


%
%
%


\end{document}